\documentclass[prl,aps,twocolumn,floatfix,10pt,longbibliography,superscriptaddress]{revtex4-1}

\usepackage{prettyref}
\newrefformat{tab}{Table \ref{#1}}
\newrefformat{fig}{Figure \ref{#1}}
\newrefformat{eqn}{Equation \ref{#1}}
\newrefformat{sec}{§\ref{#1}}

\usepackage{amsmath}
\usepackage{graphicx}
\usepackage{xcolor}

\usepackage{siunitx}

\begin{document}

\title{Structured illumination microscopy with extended axial \\ resolution through mirrored illumination}

\author{James D. Manton}
\affiliation{These authors contributed equally to the work}
\affiliation{Department of Chemical Engineering \& Biotechnology, University of Cambridge, CB3 0AS, UK}
\affiliation{Present address: MRC Laboratory of Molecular Biology, Francis Crick Avenue, Cambridge, CB2 0QH, UK}
\affiliation{Correspondence to ajdm2@cam.ac.uk}

\author{Florian Str\"ohl}
\affiliation{These authors contributed equally to the work}
\affiliation{Department of Chemical Engineering \& Biotechnology, University of Cambridge, CB3 0AS, UK}

\author{Reto Fiolka}
\affiliation{Department of Cell Biology, UT Southwestern Medical Center, 6000 Harry Hines Blvd., Dallas, Texas 75390, USA}
\affiliation{Lyda Hill Department of Bioinformatics, UT Southwestern Medical Center, 6000 Harry Hines Blvd., Dallas, Texas 75390, USA}

\author{Clemens F. Kaminski}
\affiliation{Department of Chemical Engineering \& Biotechnology, University of Cambridge, CB3 0AS, UK}

\author{Eric J. Rees}
\affiliation{Department of Chemical Engineering \& Biotechnology, University of Cambridge, CB3 0AS, UK}

\date{\today}

\begin{abstract}
Wide-field fluorescence microscopy, while much faster than confocal microscopy, suffers from a lack of optical sectioning and poor axial resolution.
3D structured illumination microscopy (SIM) has been demonstrated to provide optical sectioning and to double the achievable resolution both laterally and axially, but even with this the axial resolution is still worse than the lateral resolution of unmodified wide-field detection.
Interferometric schemes using two high numerical aperture objectives, such as 4Pi confocal and I\textsuperscript{5}S microscopy, have improved the axial resolution beyond that of the lateral, but at the cost of a significantly more complex optical setup.
Here we investigate a simpler dual-objective scheme which we propose can be easily added to an existing 3D-SIM microscope, providing lateral and axial resolutions in excess of 125 nm with conventional fluorophores.
\end{abstract}

\maketitle

\section{Introduction}
Over the past three decades, there has been a considerable effort to improve the spatial resolution of fluorescence microscopy, with the majority of advances being focussed on improving the lateral resolution.
However, in many cases, improving the axial resolution would be just as, if not more, beneficial, particularly as the axial resolution of a conventional microscope is much worse than its lateral resolution.
While both lateral and axial resolution increase with numerical aperture (NA), even an ideal single objective system that could collect over all solid angles on one side of the sample is bound to have an axial resolution at best half that of the lateral resolution.

3D structured illumination microscopy (3D-SIM) has been shown to double both lateral and axial resolution in wide-field microscopy through patterned illumination generated by the interference of three beams \cite{gustafsson_threedimensional_2008}.
As well as developments in improving temporal resolution \cite{shao_superresolution_2011,fiolka_timelapse_2012}, it has successfully been combined with dual-objective interferometric illumination and detection in I\textsuperscript{5}S microscopy to achieve an unprecedented wide-field isotropic resolution of \SI{90}{nm} \cite{shao_i5s_2008}.
Despite this impressive feat, use of this technology has been extremely limited due to the experimental difficulties in constructing and operating such a system.
Even the use of the related confocal technique of 4Pi microscopy has been limited by its complexity, with most existing systems relying on the simplest of the three 4Pi methods, Type A \cite{hell_properties_1992}.
In this article, we briefly review the theory of 3D-SIM and use this to propose a simple dual-objective scheme that does not require interferometric detection to further improve the achievable axial resolution.
In addition, by relaxing the technical constraints on the secondary objective, our scheme is compatible with low NA, high working distance objectives.

\section{3D-SIM theory}
Considering a fluorescence microscope as a linear, shift-invariant imaging system, the image data collected, \(D\), can be considered as the convolution of the fluorescence emission, \(F\), with a point spread function (PSF), \(H\):
\begin{equation}
  D = F \ast H = (S \times I) \ast H,
\end{equation}
where \(S\) is the sample fluorophore distribution and \(I\) is the illumination intensity distribution.
Alternatively, in the Fourier domain:
\begin{equation}
  \tilde{D} = \tilde{F} \times \tilde{H} = (\tilde{S} \ast \tilde{I}) \times \tilde{H},
\end{equation}
where overset tildes denote the Fourier transforms of the respective real-space functions and \(\tilde{H}\) is the optical transfer function (OTF) \cite{heintzmann_laterally_1999,gustafsson_surpassing_2000,frohn_true_2000}.

\begin{figure*}
	\centering
	\includegraphics{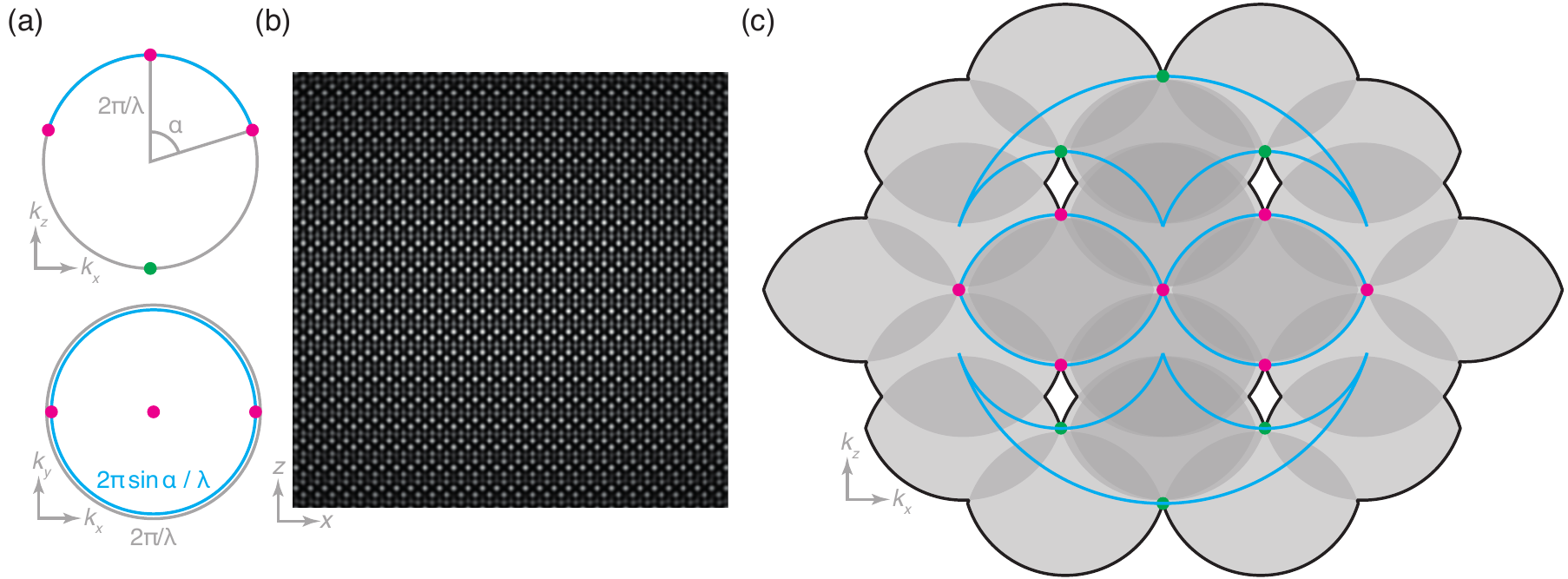}
	\caption{
		Creating an illumination profile using four mutually coherent beams.
		(a) Fourier domain illumination electric field distribution formed by four coherent beams, with those present in 3D-SIM drawn in magenta and the additional beam considered in this work drawn in green.
		Here the angular separation is the maximum possible, namely the half-angle, \(\alpha\), of the lens.
		(b) Real domain illumination intensity distribution resulting from the electric field distribution of (a).
		(c) Fourier domain illumination intensity distribution drawn as magenta and green spots (corresponding to those present in 3D-SIM and the new components considered in this work) with the overall support of the OTF drawn as a black line and each wide-field OTF contribution shown in grey.
		These illumination components can be seen to lie on the edge of the support of the 4Pi OTF (cyan lines) for this case where the angular separation is \(\alpha\).
		Note the holes within the support of the OTF near the off-axis green spots.
	}
	\label{fig:four_beams}
\end{figure*}

The illumination intensity, \(I\), is related to the electric field, \(E\), via \(I = EE^*\), where \(E^*\) is the complex conjugate of \(E\).
Hence, the Fourier domain distribution, \(\tilde{I}\), is given by the autocorrelation of \(\tilde{E}\).
In a conventional 3D-SIM system, the illumination is generated by three interfering laser beams produced by the 0 and \(\pm\)1 diffraction orders from a grating.
This means that the Fourier domain electric field amplitude distribution, \(\tilde{E}\), consists of three points on a shell of radius \(2\pi / \lambda\), spaced angularly by no more than \(\alpha\), where \(\lambda\) is the wavelength of the illumination light and \(\alpha\) is the half-angle of the objective lens (shown in magenta in \prettyref{fig:four_beams}a).
The Fourier domain distribution, \(\tilde{I}\), hence consists of the seven magenta points shown in \prettyref{fig:four_beams}c \cite{gustafsson_threedimensional_2008}.

Now, consider a modified 3D-SIM system in which a further central beam is incident upon the sample from the opposite side.
Here, the Fourier domain electric field amplitude distribution includes an additional point on the shell of radius \(2\pi / \lambda\) opposite the original central beam (green point in \prettyref{fig:four_beams}a).
This causes the Fourier domain illumination intensity distribution to gain a further six points, drawn in green in \prettyref{fig:four_beams}c.

In both cases the detection method is that of wide-field microscopy and so the overall system OTF is given by the convolution of the Fourier domain illumination intensity distribution with the conventional wide-field OTF, the bandlimit of which shown as the black line in \prettyref{fig:four_beams}c.

\section{Filling the holes in the OTF}
Taking the 13-point illumination structure and convolving with the conventional wide-field OTF produces a system OTF with larger axial support but, crucially, contains holes (grey fill in \prettyref{fig:four_beams}c).
In order for all sample frequencies within the maximum extent of the OTF to be accurately recorded, these holes must be filled.
This cannot be achieved by merely changing the angular spread of the outer points of the illumination and so alternative approaches must be sought.
Here, we consider three approaches to achieve this.
In the following discussion, examples are shown for an oil-immersion 1.45 NA lens operating in a nominal refractive index of 1.518.
Unless otherwise stated, illumination is at \SI{488}{nm} with fluorescence emission detected at \SI{510}{nm}.

\subsection{Illuminating at two wavelengths}
Consider the effect of illuminating with two different wavelengths.
While the overall structure of the illumination will be similar in each case, the wavelength difference will cause the distance from the origin of each Fourier domain illumination component to differ (compare purple and green points in \prettyref{fig:filling_holes}a).
By selecting an appropriate combination of wavelengths, such as \SI{445}{nm} and \SI{561}{nm} illumination with \SI{610}{nm} detection, the holes for each wavelength can be filled with information obtained using the other illumination wavelength.
With this, nine phase steps rather than five are required, slowing acquisition speed.
In addition, the structure of interest now needs to be labelled with fluorophores with a broad excitation spectrum, such as quantum dots, complicating multicolour imaging of different structures.
Alternatively, two different fluorophores with well-separated excitation and emission spectra can be used, with two sets of five-phase data acquired and fused, at the expense of a further reduction in acquisition speed.

\subsection{Recording data with multiple pattern orientations}
So far we have only considered a two-dimensional system operating in the \(xz\)-plane (where \(z\) is the optical axis).
In a true three-dimensional system the wide-field OTF is not the bi-lobed object depicted in cyan in \prettyref{fig:four_beams}c, but the toroidal solid of revolution given by that cross-section \cite{frieden_optical_1967}.
Hence, by recording data with three pattern orientations spaced by \SI{120}{\degree}, the holes in one orientation are filled by contributions from the next (as shown in \prettyref{fig:filling_holes}b).
However, this precludes acquisitions using just one orientation, which may be desired for reasons of speed when enhancing the axial resolution is more important than an isotropic lateral resolution enhancement.

\subsection{Illuminating with multiple modes}
Finally, consider illuminating not with single modes in the pupil, but with multiple modes over a significant area of the pupil.
Each order generated by the grating produces an image of the source in the pupil.
For appropriately spatially incoherent illumination, each point within each source image will be incoherent with all other points within the same image, but coherent with the corresponding points in the other source images.

\begin{figure*}
	\centering
	\includegraphics{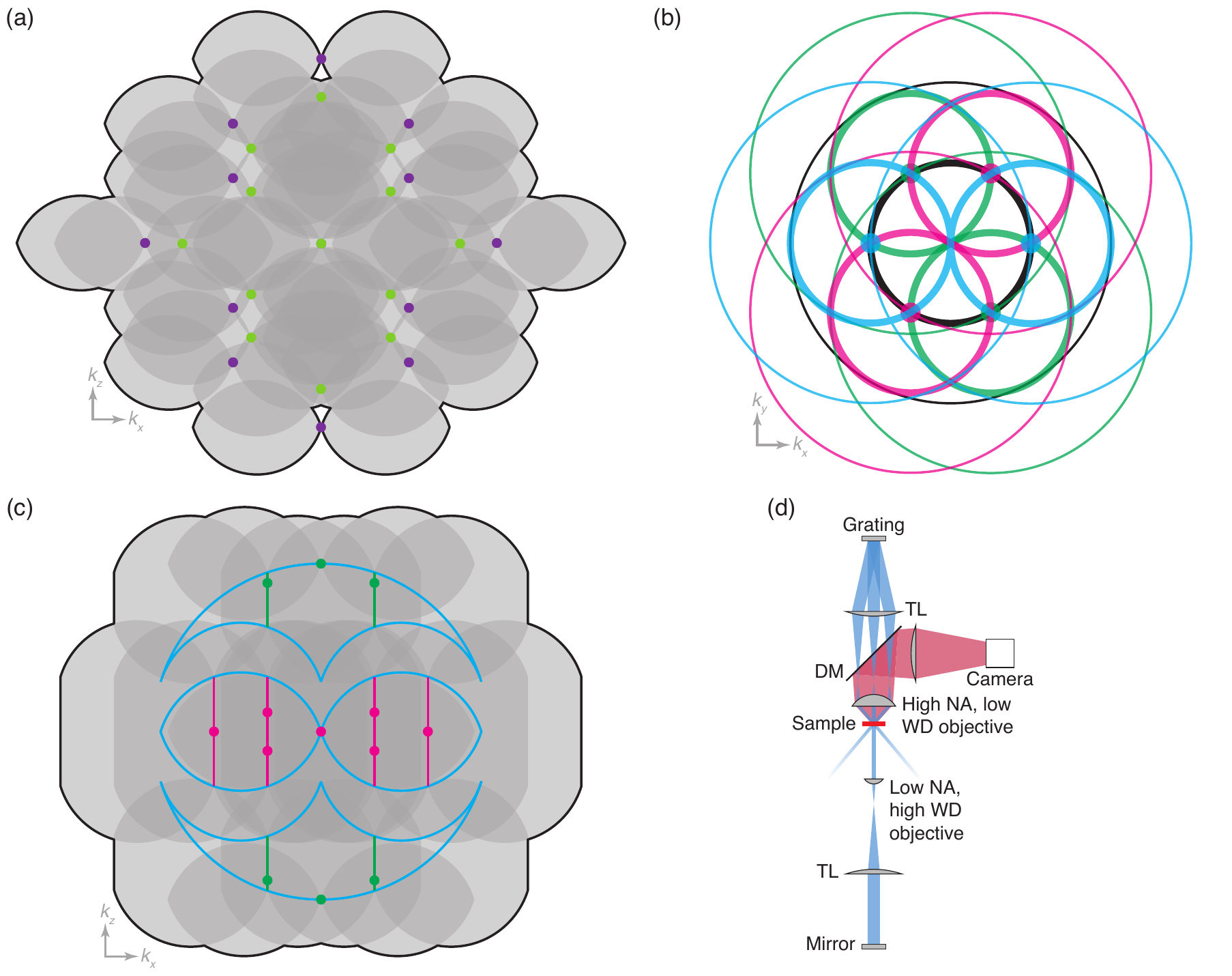}
	\caption{
		Strategies for filling the holes in the OTF.
		(a) Overall effective OTF (support shown as black line) for illumination at \SI{445}{nm} (purple spots) and \SI{561}{nm} (yellow-green spots) with fluorescence detection at \SI{610}{nm} (gray OTFs).
		Here the holes are filled due to the difference in \(k\)-vector length, but two new ones are created along \(k_z\). These can easily be removed by apodisation at the cost of slightly reduced resolution.
		(b) The effect of considering multiple orientations on the \(k_y\)--\(k_x\) coverage.
		The ever-present zero order wide-field detection OTF component is shown in black, with the thin line denoting the edge of the OTF support, where the axial extent is lowest, and the thick black line denoting the region of the OTF where the axial extent is highest.
		The first order contributions from each orientation are shown in cyan, magenta and green, with the holes in the OTF support from each orientation being shown by the large filled circles.
		As can be seen from the overlap of thick lines with circles, each hole is effectively filled in by contributions from the next orientation.
		(c) The effect of multimode illumination with a pupil fraction of \(1/3\).
		Considering the centre of each illumination image created by the grating leads to the Fourier domain illumination distribution shown as the magenta and green spots.
		Summing over all such point tetrads causes an axial broadening of this distribution, shown as the magenta and green lines.
		While no contributions from outside the 4Pi OTF exist, the shifting of contributions within this ensures a good overlap and hence no holes in the overall OTF support.
		(d) Schematic of the proposed experimental setup in which a conventional 3D-SIM microscope is augmented with a low numerical aperture, high working distance objective on the other side of the sample, paired with a tube lens and mirror to reflect just the central beam.
	}
	\label{fig:filling_holes}
\end{figure*}

As the objective lens obeys the Abbe sine condition, points in the pupil are projected orthographically onto the shell of radius \(2\pi/\lambda\) when considering the Fourier domain electric field amplitude distribution.
Hence, for each set of source point images, the lateral components of the Fourier domain illumination intensity distribution will not change while the axial ones will.
The end effect, after summing over all points within the source images, is an illumination intensity distribution in which each of the original points (except for ones along \(k_z\)) have been broadened axially but not laterally (green and magenta lines in \prettyref{fig:filling_holes}c and Visualisation 1).

The outermost components are broadened to the full axial extent of the wide-field OTF at that lateral frequency, while the inner components are broadened half as much.
For a pupil fraction of \(1/3\) (the maximum possible before source images overlap), corresponding to a maximum lateral illumination frequency of \(2/3\) that of the maximum supported by the lens, the effect is such that all components are broadened to the full extent of the local component of the 4Pi OTF (compare green and magenta lines with cyan 4Pi OTF extent in \prettyref{fig:filling_holes}c).

While the exact absolute extent of the axial broadening and overlap is dependent on the exact numerical aperture, refractive index of immersion, \(n\), and pupil fraction used, a pupil fraction of \(1/3\) is more than sufficient to fill all the holes for a 1.45 numerical aperture oil immersion objective (\(n = 1.518\)) as demonstrated in \prettyref{fig:filling_holes}c.
While this approach does not require any changes to the sample labelling and fills the holes with single-orientation reconstruction procedures, the maximum possible lateral resolution enhancement is decreased as the illumination pattern is formed by beams spaced by a lower numerical aperture \cite{gustafsson_threedimensional_2008}.
In practice, using this method with a small pupil fraction and combining data from multiple pattern orientations should provide an improved lateral resolution enhancement while still ensuring good Fourier domain coverage within the full support of the OTF.

\section{Considerations for an experimental realisation}
The required fourth beam can be provided via a relatively simple modification to an existing 3D-SIM system.
By placing a low numerical aperture, high working distance objective on the opposite side of the sample, the central beam can be captured while the outer beams are left to escape.
A tube lens and mirror placed in the image plane reflects this central beam, sending a fourth beam into the sample as required (see \prettyref{fig:filling_holes}d).

As we require this reflected beam to interfere with the three incident from the high NA objective side of the sample, the coherence length of the laser must be greater than the \(\sim\)\SI{800}{mm} extra path length experienced by the reflected beam.
Hence, for a \SI{488}{nm} laser line, the frequency bandwidth must be less than \(\sim\)\SI{100}{MHz}.
While laser sources conventionally used for SIM may not have a sufficiently small bandwidth, appropriate single frequency lasers are readily available from commercial sources.

In order for the central beam to be fully captured by the low NA objective, the NA of this lens must be greater than the product of the NA of the high NA objective with the illumination pupil fraction.
Objective lenses with NAs of 0.5 and working distances greater than \SI{1}{mm} are commercially available and would be suitable for capturing the central beam from a 1.45 NA primary objective, even if the pupil fraction is the maximum possible (\(1/3\)).

If the multimode approach is to be used, illumination can be provided by laser light coupled into a shaken multimode fibre as this well-approximates a spatially incoherent source over sufficiently long times (typically of the order of less than a millisecond) \cite{fiolka_timelapse_2012}.
As the diameter of the objective pupil is given by \(2f \times \textrm{NA}\), where \(f\) is the back focal length and \(\textrm{NA}\) is the numerical aperture, a \(1/3\) pupil fraction would correspond to a \(\sim\)20-fold magnification for a \SI{100}{\micro m} core fibre and a 100\(\times\) objective designed for a \SI{200}{mm} tube length.
Such fibres are commonly available with an NA of 0.22, ensuring minimal loss of light through the relay optics from fibre tip to pupil image while still satisfying the conditions for proper demagnification of the grating onto the sample.

\section{Discussion}
We have investigated the possibility of further axial resolution enhancement in 3D-SIM by reflecting the central beam and have provided three methods for countering the existance of holes in the OTF support.
While further resolution enhancements would be possible by using another high NA lens as the secondary objective to provide a further two illumination beams, and by using interferometric detection as in I\textsuperscript{5}M and I\textsuperscript{5}S microscopy \cite{gustafsson_i5m_1999,shao_i5s_2008}, the comparatively simple experimental setup proposed here achieves much of the performance of such a system while avoiding the significant complexity of interferometric detection.

For an oil immersion objective of 1.45 NA, working at \SI{488}{nm} excitation and \SI{510}{nm} detection, a \SI{125}{nm} axial resolution should be achievable using the multimode illumination scheme presented, while still providing \SI{105}{nm} lateral resolution.
While this scheme is incompatible with normal 1.2 NA water objectives, a commercially available 1.27 NA water objective has a similar collection angle to a 1.45 NA oil objective and so this scheme has the potential to be compatible with live-cell imaging.
If increased lateral resolution is not important, a truly isotropic 3D resolution of around \SI{200}{nm} could be provided at a acquition rate only 15-fold slower than wide-field microscopy (with five phases and three-fold finer axial steps, to satisfy Nyquist sampling).
While the results presented here are based on geometrical considerations, supporting numerical simulations are provided in Supplementary Code Files 1--3.

\section{Funding Information}
We acknowledge funding from the UK Engineering \& Physical Sciences Research Council (EP/L015889/1, EP/H018301/1 and EP/G037221/1), the Wellcome Trust (3-3249/Z/16/Z and 089703/Z/09/Z), the UK Medical Research Council (MR/K015850/1 and MR/K02292X/1), MedImmune, and Infinitus (China) Ltd.

\bibliography{references}

\end{document}